\documentclass{aa}
\input epsf
\usepackage{graphics}

\def\Msun{M_{\sun}}
\def\Lsun{L_{\sun}}

\def\Mstar{M_{\star}}
\def\Lstar{L_{\star}}

\def\Tstar{T_{\star}}

\def\CzuO{\varepsilon_{\rm C}/\varepsilon_{\rm O}}
\def\dup{\Delta u_{\rm p}}

\def\kd{\kappa_{\rm d}}
\def\fcond{f_{\rm c}}
\def\uex{\langle u \rangle}

\def\lra#1{\langle #1 \rangle}

\def\msyr{{\rm \Msun / yr}}

\def\Jt{{\it J\"ager1000}}   
\def\Jf{{\it J\"ager400}}    
\def\MA{{\it Maron}}
\def\RO{{\it Rouleau}}
\def\ZU{{\it Zubko}}
\def\PR{{\it Preibisch}}

\begin{document}

   \thesaurus{07     
              (02.08.1;  
               02.18.7;  
               08.01.3;  
               08.03.1;  
               08.16.4  
               )} %
   \title{Optical properties of carbon grains: Influence
            on dynamical models of AGB stars}

   \subtitle{}

   \author{Anja C.\,Andersen
           \inst{1}
           \and
           Rita Loidl\inst{2}
           \and
           Susanne H\"{o}fner\inst{2,3}
          }

   \offprints{A.\,C.\,Andersen}

   \institute{
              Niels Bohr Institute, Astronomical Observatory,
              Juliane Maries Vej 30, DK-2100 Copenhagen, Denmark\\
              email: anja@astro.ku.dk
   \and
              Institute for Astronomy, University of Vienna,
              T{\"u}rkenschanzstra{\ss}e 17, A-1180 Vienna, Austria\\
              email: loidl@astro.univie.ac.at
   \and
              Uppsala Astronomical Observatory, Box 515, SE-75120 Uppsala,
              Sweden \\
              email: hoefner@astro.uu.se}

   \date{Received 14 April 1999 / Accepted 7 July 1999}

   \maketitle

   \begin{abstract}

For amorphous carbon several laboratory extinction 
data are available, which show quite a wide range of 
differences due to the structural complexity of this material. 
We have calculated self-consistent dynamic models of circumstellar 
dust-shells around carbon-rich asymptotic giant branch stars,
based on a number of these data sets.
The structure and the wind properties of the dynamical models
are directly influenced by the different types of amorphous carbon.
In our test models the mass loss is not severely dependent on
the difference in the optical properties of the dust, but the influence
on the degree of condensation and the final outflow velocity is considerable.
Furthermore, the spectral energy distributions and colours resulting from
the different data show a much wider spread than the variations within the 
models due to the variability of the star.
Silicon carbide was also considered in the radiative transfer calculations
to test its influence on the spectral energy distribution.

   \end{abstract}

%

\section{Introduction}

Asymptotic giant branch (AGB) stars show large amplitude 
pulsations with periods of about 100 to 1000 days.
The pulsation creates strong shock waves in the
stellar atmos\-phere, causing a levitation of the outer layers. This cool
and relatively dense environment provides favourable conditions for the
formation of molecules and dust grains. Dust grains play an important 
role for the heavy mass loss, which influences
the further evolution of the star.

Condensation and evaporation of dust in envelopes of pulsating stars
must be treated as a time-dependent process since the time scales for
condensation and evaporation are comparable to variations of the
thermodynamic conditions in the stellar envelope.  The radiation pressure on
newly formed dust grains can enhance or even create shock
waves leading to more or less pronounced discrete dust shells in the expanding
circumstellar flow (e.g.\ Fleischer et al.\ \cite{Fleischer91},
\cite{Fleischer92}; H\"ofner et al.\ \cite{Hoefner95}; H\"ofner \& Dorfi
\cite{Hoefner97}). Since a significant part of the dust grains transferred to 
interstellar space
are produced in the atmosphere of these old luminous stars (Sedlmayr
\cite{Sedlmayr94}) an understanding of the nature of mass loss of these
long-period variables is crucial for the general understanding of dust
in space.

Modelling circumstellar envelopes requires knowledge of the 
absorption properties of the different types of grains over 
the relevant part of the electromagnetic spectrum.
For this the optical properties of the corresponding dust material are needed.
Amorphous carbon is a very good candidate as the most common type of carbon
grains present in circumstellar envelopes, since the far-infrared data of
late-type stars show a spectral index as expected for a very
disordered two-dimensional material like amorphous carbon 
(Huffman \cite{Huffman88}). 

Silicon carbide (SiC) grains seem to be another
important component of the dust in circumstellar envelopes. 
While amorphous carbon could explain the
continuum emission, SiC particles could be responsible for the 11.3 $\mu$m
band observed in many C-rich objects.

 In this paper, we present self-consistent dynamical
models of circumstellar dust
shells calculated with selected laboratory amorphous carbon data.
Based on these mo\-dels we have performed radiative transfer calculations
for pure amorphous carbon and in some cases also inclu\-ding SiC dust.
In Sect.\,\ref{opaci} the used amorphous carbon data are described. 
The influence on the model structure is described in 
Sect.\,\ref{s:dynmod} and the resulting spectral appearance
is discussed in Sect.\,\ref{sed}.

\section{Optical properties of dust}\label{mietheory}

The two sets of quantities that are used to describe optical
properties of solids 
are the real and imaginary parts of the complex refractive index
$m = n + ik$ and the real and imaginary parts of the complex dielectric
function (or relative permittivity) $\epsilon$ = $\epsilon'$ + $i
\epsilon''$.  These two sets of quantities are not independent, the
complex dielectric function $\epsilon$ is related to the complex refractive
index, $m$, by $\epsilon' = n^{2} - k^{2}$ and $\epsilon'' = 2 nk$,
when the material is assumed to be non-magnetic ($\mu = \mu_{0}$).
Reflection and transmission by bulk media are best described using
the complex refractive index, $m$, whereas absorption and scattering
by particles which are small compared with the wavelength are best
described by the complex dielectric function, $\epsilon$.

The problem of evaluating the expected spectral dependence of extinction
for a given grain model (i.e.\ assumed composition and size
distribution) is essentially that of evaluating the extinction
efficiency Q$_{\rm ext}$. It is the sum of 
corresponding quantities for absorption and scattering;
${\rm Q}_{\rm ext} ={\rm Q}_{\rm abs}  +{\rm Q}_{\rm sca}$.
These efficiencies are functions of two quantities; a dimension-less size
parameter $x = 2 \pi a/ \lambda$ (where $a =$ the grain radius and 
$\lambda =$ the wavelength) and a composition parameter, the complex refractive 
index $m$ of the grain material.
Q$_{\rm abs}$ and Q$_{\rm sca}$ can therefore be calculated 
from the complex refractive index using Mie theory
for any assumed grain model.
The resulting values of total extinction can be compared with observational
data.  

A limit case within the Mie theory is the Rayleigh approximation for
spherical particles. This approximation is
valid when the grains are small compared to the
wavelength, $x = 2 \pi a / \lambda << 1$
and in the limit of zero phase shift in the particle ($|m|x << 1$).
In the Rayleigh approximation the extinction by a sphere in vacuum is
given as:
\begin{equation}
\frac{Q_{ext}}{a} =  \frac{8 \pi}{\lambda} {\rm Im} \{ 
                     \frac{m^{2} -1}{m^{2} +2} \}.
\end{equation}

\subsection{Measuring methods of optical properties}

A proper application of Mie theory to experimental data requires
that the samples are prepared such that the particles are quite
small (usually sub-micrometer), well isolated from one another, and that the
total mass of particles is accurately known. 

In order to obtain single isolated homogeneous particles, 
the grains are often dispersed in a solid matrix. 
Small quantities of 
sample are mixed throughly with the powdered matrix material 
e.g.\ KBr or CsI.  The matrix is pressed into a
pellet which has a bulk transparency in the desired wavelength region. 
Some of the problems with this technique are that there is a tendency
for the sample to clump along the outside rim of the large matrix grains
and that the introduction of a matrix, which has a refractive index different
from vacuum, might influence the band shape and profile. This matrix effect
can be a problem for comparisons of laboratory measurements
with astronomical spectra (Papoular et al.\ \cite{Papoular98}; Mutschke et al.\
\cite{Mutschke99}).

By measuring the sample on a substrate (e.g.\ quartz, KBr, Si or NaCl)
using e.g.\ an infrared microscope,
the matrix effect can nearly be avoided since the sample is almost fully 
surrounded by a gas (e.g.\ air, Ar or He). 
But the amount of material in the microscopic aperture remains unknown,
which is an important disadvantage of this method. 
Therefore, these measurements are not quantitative but they reveal the shape 
of the spectrum nearly without a matrix effect (Mutschke et al.\ 
\cite{Mutschke99}).

A major problem of both methods is clustering 
of the grain samples either during the production of the partic\-les or
within the matrix or on the substrate.
Clustering can cause a dramatic difference in the optical properties
(Huffman \cite{Huffman88}).
A way to avoid this problem is to perform the optical measurements
on a polished bulk sample. For the determination of
both $n$ and $k$ two or 
more measurements on bulk samples are required. 
This might be done either by a
transmission and a reflection measurement, or by  two reflectance measurements
determinations at different angles or with different polarisations.  
Since the real part, $n$,
of the refractive index, $m$, is determined by the phase velocity and the
imaginary part, $k$, by the absorption, a transmission measurement
easily fixes $k$. 
The Kramers-Kronig relations can be applied in order to obtain the 
optical constants for grain measurements. 
The real part of the refractive index can be expressed as an 
integral of the imaginary part (see e.g.\ Bohren \& Huffman \cite{Bohren83}).

\section{Carbon grains}\label{opaci}

While carbon is expected to constitute a major fraction of the circumstellar 
dust in carbon stars, its exact form is still unclear. 
Carbon has the unique property that the atoms can form three different
types of bonds through sp$^{1}$, sp$^{2}$ (graphite) and sp$^{3}$ (diamond)
hybridization.

A number of observations of late-type stars contradict the presence of 
graphite as the dominant dust type (e.g. Campbell et al.\ \cite{Campbell76}; 
Sopka et al.\ \cite{Sopka85}; Martin \& Rogers \cite{Martin87}; 
G\"urtler et al.\ \cite{Gurtler96}). The far-infrared (FIR)
data of late-type stars generally show a dust emissivity law of 
$Q(\lambda) \sim \lambda ^{- \beta}$ with a spectral index of $\beta
\approx 1$. While graphite grains have a FIR emission proportional to
$\lambda ^{-2}$ (Draine \& Lee \cite{Drain84}),  
a $\lambda ^{-1}$ behaviour can be
expected in a very disordered two-dimensional material like amorphous
material (Huffman \cite{Huffman88}; J\"ager et al.\ \cite{Jager98}).

Amorphous carbon grains therefore seem to be a very good candidate as the common
type of carbon grains present in circumstellar envelopes. Another possibility 
could be small diamond grains. Presolar diamond grains have been identified
from primitive (unaltered) meteorites (carbonaceous chondrites) and are the
most abundant (500 ppm) of the presolar grains discovered so far (Lewis et
al.\ \cite{Lewis87}). At least 3\% of the total amount of carbon present at the
formation of the Solar System was in the form of diamonds (Huss \& Lewis
\cite{Huss94}). The place of origin of the presolar diamonds is still 
unknown, but since they can only have formed under reducing conditions 
J{\o}rgensen (\cite{Jorgensen88}) has suggested C-rich AGB stars as
the place of formation of the majority of the presolar diamond grains.

It has been suggested by Kr\"uger et al.\ (\cite{Kruger96}) 
that the surface growth processes on carbonaceous seed particles 
in circumstellar envelopes will take place at sp$^{3}$ bonded 
carbon atoms rather than at
sp$^{2}$ bonded ones, which suggests that the grain material 
formed in circumstellar envelopes could be amorphous-diamond like carbon.
Presolar
diamonds extracted from meteorites have a median grain size of 
about 2 nm (Fraundorf et al.\ 
\cite{Fraundorf89}), meaning that each diamond contains a few hundred to
a few thousand carbon atoms.  
The presolar diamonds therefore seem to 
actually consist of a mixture of diamond (core) and hydrogenate amorphous
carbon (surface) having about 0.46 the volume fraction of pure diamond
(Bernatowicz et al.\ \cite{Bernatowicz90}).

Several spectra of presolar diamonds from various meteorites have been
published (Lewis et al.\ \cite{Lewis89}; Colangeli et al.\ \cite{Colangeli95};
Mutschke et al.\ \cite{Mutschke95}; Hill et al.\ \cite{Hill97}; Andersen et al.\
\cite{Andersen98}; Braatz et al., submitted to Meteorit.\ Planet.\ Sci.) 
and even though a number
of artifacts tends to be present in all the spectra, the general trend is that
the presolar diamonds have an absorption coefficient that is twice that of
pure diamond and a factor of a hundred less than the ``diamond-like''
amorphous carbon of J\"ager et al.\ (\cite{Jager98}). 

There exists a wide variety of possible amorphous carbon grain types,
which fall in between the categories ``diamond-like'' and 
``graphite-like'' amorphous carbon.
We have calculated dynamical models using various laboratory data of 
amorphous carbon to determine the possible influence of these different 
grain types on the structure and the wind properties of C-rich AGB star
models. 

\subsection{Laboratory measurements of amorphous carbon}\label{lab}

Laboratory conditions are far from the actual space conditions where
grains are produced or processed, but experiments in which physical
and chemical parameters are controlled and monitored do give 
the option of selecting materials which may match the astronomical observations.
When choosing which amorphous carbon data to use one is faced with the
fact that due to the various processes used in the sample preparation, 
differences often appear between the measurements of various authors.
Another major problem is that the optical properties of amorphous carbon
are most often obtained by different techniques in different wavelength
regions. 
Extinction measurements of sub-micron-sized particles is the most common
technique in the infrared. 
In the visible and ultraviolet,
reflecti\-vity and transmission measurements are often obtained on bulk samples.

\begin{figure}
 \centering 
 \leavevmode 
 \epsfxsize=1.0 
 \columnwidth
 \epsfbox{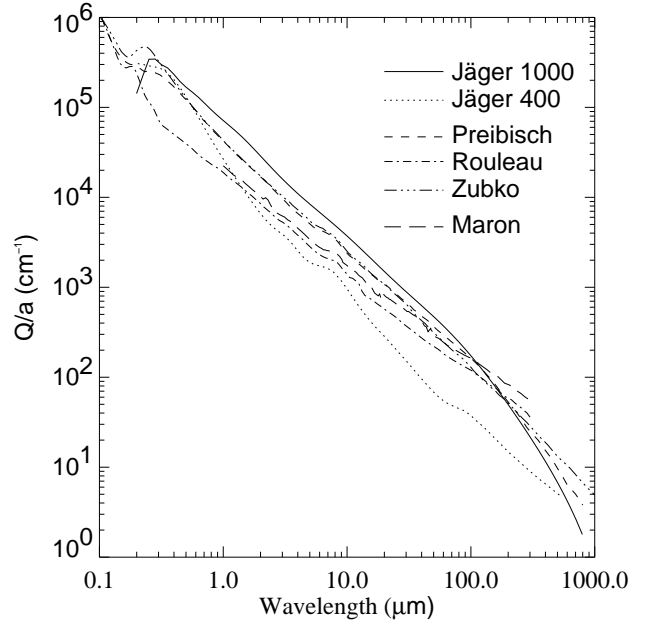}
 \caption[]{Calculated extinction efficiency of amorphous carbon
from optical constants published by Maron (\cite{Maron90}), Rouleau \& Martin
(\cite{Rouleau91}), Preibisch et al.\ (\cite{Preibisch93}), Zubko et al.\ 
(\cite{Zubko96}) and J\"ager et al.\ (\cite{Jager98}), see Table \ref{opacities}
for annotations.
The extinction efficiency factors were calculated from the optical
constants ($n$ and $k$) in the Rayleigh approximation for spheres.}
 \label{nk}
\end{figure}

Bussoletti et al.\ (\cite{Bussoletti87a}) have  determined the extinction 
efficiencies for
various types of sub-micron amorphous carbon particles
and spectroscopically analysed them in the wavelength range 1000 {\AA} --
300 $\mu$m. In their paper they present an updated version of the
data already published from 2000 {\AA} to 40 $\mu$m (Borghesi et al.\ 
\cite{Borghesi83}, \cite{Borghesi85a}) and new data obtained in the UV/vis 
(1000 -- 3000 {\AA}) and in
the FIR (35 -- 300 $\mu$m). The sub-micron amorphous carbon grains were
obtained by means of two methods: (1) striking an arc between two
amorphous carbon electrodes in a controlled Ar atmosphere at different 
pressures (samples AC1, AC2 and AC3; where the numbers 
refer to different accumulation distances from the arc discharge); (2) burning 
hydrocarbons (benzene and xylene) in air at room pressure (samples BE and XY).
The smoke was collected on quartz substrates. For the UV/vis spectroscopy
the quartz substrates on which the particles had been collected were used 
directly while the dust was scrapped from the substrate and embedded in
KBr pellets for the IR spectroscopy. Bussoletti et al.\ 
(\cite{Bussoletti87b}) suggest 
that the extinction efficiencies, $Q_{\rm ext}/a$, for the AC samples
should be 
corrected by a factor of 5 due to an experimental underestimation of the pellet
density. This correction gives an agreement with the data 
by Koike et al.\ (\cite{Koike80}). 

Colangeli et al.\ (\cite{Colangeli95}) measured the extinction efficiency in the range
40 nm -- 2 mm. They produced three different samples; two by arc discharge
between amorphous carbon electrodes in Ar and H$_{2}$ atmospheres at 10 mbar
(sample ACAR and ACH2 respectively) and one by burning benzene in air (sample
BE). The samples were deposited onto different substrates for the UV/vis
measurements, while in the IR the samples where prepared both on a substrate
and by being embedded in  KBr/CsI pellets and 
for the FIR measurements the samples
were embedded in polyethylene pellets. These different but overlapping methods
gave the possibility of evaluating the difference as a result of embedding
the samples in a matrix or by ha\-ving them on a substrate.
Colangeli et al.\ (\cite{Colangeli95}) found that embedding the samples
in a matrix introduces a systematic error (the matrix effect) while the spectra
obtained for grains deposited onto a substrate did not suffer from any matrix
effect detectable within the accuracy available in the experiment.
Therefore the FIR data were corrected for the extinction offset introduced 
by the matrix.

J\"ager et al.\ (\cite{Jager98}) produced structurally different carbon
materials by pyrolizing cellulose materials at different temperatures
(400$^{\circ}$\,C, 600$^{\circ}$\,C, 800$^{\circ}$\,C and 1000$^{\circ}$\,C),
and characterised them in great detail.
These materials have increasing sp$^{2}$/sp$^{3}$ ratios making the
amorphous carbon pyro\-lysed at 400$^{\circ}$\,C the most "diamond-like" 
with the lowest sp$^{2}$/sp$^{3}$ ratio while the amorphous carbon pyro\-lysed at
1000$^{\circ}$\,C is more "graphite-like" with the highest sp$^{2}$/sp$^{3}$
ratio. The
pyro\-lysed carbon samples were embedded in epoxy resin and reflectance of
the samples was measured in the range 200 nm to 500 $\mu$m, making
this the first consistent laboratory measurement of amorphous carbon over
the whole spectral range relevant for radiative transfer calculations
of C-rich AGB stars.  
From the reflectance spectra the complex refractive index, $m$,
was derived by the Lorentz oscillator method (see e.g.\ Bohren \& Huffman
1983, Chap.\ 9).
There is a significant difference between the two low temperature 
(400$^{\circ}$\,C and 600$^{\circ}$\,C) and the two high temperature
samples (800$^{\circ}$\,C and 1000$^{\circ}$\,C).  
The latter two behave very similar to glassy carbon. 

In contrast to grain measurements, the bulk samples by 
J\"ager et al.\ (\cite{Jager98}) 
give the possibility of investiga\-ting the difference between the influence 
of the internal structure of amorphous material and the morphology of the
carbon grains. These two properties can be separated out due to the
careful investigation of the internal structures of the four samples and
the range of material properties that these four amorphous carbon samples
span (from "diamond-like" to "graphite-like").

\begin{table*}
\caption[]{Comparison of the different laboratory data and a list
of authors who have obtained optical constants from these data.}
\label{opacities}
\begin{tabular}{|l|c|c|c|c|l|} \hline 
Reference & Material & $\rho$ & Wavelength & Designation &  Comments \\ 
  & name & (g/cm$^{3}$) & interval ($\mu$m)& in this paper & \\ \hline \hline
Bussoletti et al.\ (1987a) & AC2 & 1.85 & 0.1 -- 300 & Bussoletti~AC-2 & Arc discharge \\ 
Bussoletti et al.\ (1987a) & BE  & 1.81$^{1}$ & 0.2 -- 300 & Bussoletti~BE & burning benzene \\ 
Bussoletti et al.\ (1987a) & XY  &  & 0.2 -- 300 & Bussoletti~XY & burning Xylene \\ 
Colangeli et al.\ (1995) & ACAR & 1.87 & 0.04 -- 2000 & Colangeli~AC & arc discharge in Ar \\
Colangeli et al.\ (1995) & ACH2 &  & 0.04 -- 950 & Colangeli~ACH2 & arc discharge in H$_{2}$ \\
Colangeli et al.\ (1995) & BE &  & 0.05 -- 2000 & Colangeli~BE & burning benzene \\
J\"{a}ger et al.\ (1998) & cel400 & 1.435 & 0.02 -- 500 & J\"{a}ger~400 & most diamond-like \\ J\"{a}ger et al.\ (1998) & cel600 & 1.670 & 0.02 -- 500 & J\"{a}ger~600 &  \\
J\"{a}ger et al.\ (1998) & cel800 & 1.843 & 0.02 -- 500 & J\"{a}ger~800 & \\
J\"{a}ger et al.\ (1998) & cel1000 & 1.988 & 0.02 -- 500 & J\"{a}ger~1000 & most graphite-like \\ \hline \hline
Maron (1990) & AC2 & & & Maron & n \& k from Bussoletti et al.\ (1987a) \\
Rouleau \& Martin (1991) & AC2 & & & Rouleau & n \& k from Bussoletti et al.\ (1987a) \\
Preibisch et al.\ (1993) & BE & & & Preibisch & n \& k from Bussoletti et al.\ (1987a) \\
Zubko et al.\ (1996) & ACAR & & & Zubko & n \& k from Colangeli et al.\ (1995) \\ \hline
\end{tabular}
$^{1}$Given in Rouleau \& Martin (1991).
\end{table*}
 
\subsection{Calculated optical properties of amorphous carbon}

Several authors have used the data of
Bussoletti et al.\ (\cite{Bussoletti87a}) and 
Colangeli et al.\ (\cite{Colangeli95}) to obtain the optical constants 
of amorphous carbon grains. 

Maron (\cite{Maron90}) used the extinction efficiencies of
Bussoletti et al.\ (\cite{Bussoletti87a}) 
(sample AC2) to derive the optical constants 
(n and k) by estimating
the complex permittivity by a combination of the measured absorption 
efficiencies, dispersion formulae and Kramers-Kronig relation.
The reason for performing these calculations is that the optical constants
are needed for modelling emission properties of grains containing various 
allotropic carbons or having different sizes. 
Maron (\cite{Maron90}) is of the opinion that the differences between the 
primary extinction efficiencies obtained by Bussoletti et al.\
(\cite{Bussoletti87a}) and 
Koike et al. (\cite{Koike80}) are real and caused rather by the use of 
different electrodes
(amorphous carbon and graphite, respectively) than by an underestimation of
the pellet column density as suggested by Bussoletti et al.\ 
(\cite{Bussoletti87b}).
Therefore he did not introduce the correction suggested by Bussoletti et al.\
(\cite{Bussoletti87b}).

Rouleau \& Martin (\cite{Rouleau91}) used the AC2 and BE data from 
Bussoletti et al.\ (\cite{Bussoletti87a}) 
to produce synthetic optical constants ($n$ and $k$) which 
satisfy the Kramers-Kronig relations and highlight the effects of assuming
various shape distributions and fractal clusters. 
One of the complications in determining these optical properties of 
amorphous carbon material was that in the infrared the extinction 
measurements were done on a sample of sub-micron-sized particles, while in
the visible and ultraviolet the optical constants were obtained by 
measurements of reflectivity and transmission or by electron energy loss
spectroscopy on bulk samples. These diverse measurements were used
to produce synthetic optical constants which satisfied the Kramers-Kronig 
relations. 

Preibisch et al.\ (\cite{Preibisch93}) used the BE sample from 
Bussoletti et al.\ (\cite{Bussoletti87a})
between 0.1--300 $\mu$m and the data of Blanco et al.\ 
(\cite{Blanco91}) between 40--700 $\mu$m,
using the same technique as used by Rouleau \& Martin 
(\cite{Rouleau91}) for deriving optical
constants taking shape and clustering effects into account. Preibisch et al.\
(\cite{Preibisch93})
extend the available optical constants on the basis of the measurements
of Blanco et al.\ (\cite{Blanco91}). With these they determine the opacities of
core-mantle-particles with varying mantle thickness and pollution.

Zubko et al.\ (\cite{Zubko96})  
used the extinction efficiencies obtained by Colangeli et al.\
(\cite{Colangeli95})
to derive the optical constants (n and k) also by use of the
Kramers-Kronig approach. 
These data were used to evaluate the possible shapes of the amorphous 
carbon grains in space and the possible clustering of the particles.
 
In this study we have used the derived optical constants of Maron (\cite{Maron90}),
Rouleau \& Martin (\cite{Rouleau91}), Preibisch et al.\ (\cite{Preibisch93}), 
Zubko et al.\ (\cite{Zubko96}) and J\"ager et al.\ (\cite{Jager98}), see
Table \ref{opacities} for details. The extinction efficiency data presented 
in this paper were calculated in the Rayleigh approximation for spheres.

\subsection{The nature of silicon carbide}\label{SiC}

Thermodynamic equilibrium calculations performed by Friedemann (1969a,b)
\nocite{Friedemann69a} \nocite{Friedemann69b} and Gilman (\cite{Gilman69}) 
suggested that SiC particles
can form in the mass outflow of C-rich AGB stars. 
The observations performed
by Hackwell (\cite{Hackwell72}) and Treffers \& Cohen  
(\cite{Treffers74}) presented the first empirical 
evi\-dence for the presence of SiC particles in stellar atmos\-pheres.  
A broad infrared emission
feature seen in the spectra of many carbon stars, peaking between 11.0
and 11.5 $\mu$m is therefore attributed to solid SiC particles and SiC
is believed to be a significant constituent of the dust around carbon
stars.

An ultimate proof for the formation of SiC grains in C-rich stellar
atmospheres was the detection of isotropically anomalous SiC grains in
primitive meteorites (Bernato\-wicz et al.\ \cite{Bernatowicz87}). 
Based on isotopic measurements of the
major and trace elements in the SiC grains and on models of stellar
nucleosynthesis, it is established that a majority of the presolar SiC
grains has their origin in the atmos\-pheres of late-type C-rich
stars (Gallino et al.\ \cite{Gallino90}, 1994; Hoppe et al.\
\cite{Hoppe94}). For recent reviews
see, e.g., Anders \& Zinner (\cite{Anders93}), Ott (\cite{Ott93}) and 
Hoppe \& Ott (\cite{Hoppe97}). 

Detailed laboratory investigations on the infrared spectrum of SiC
have been presented by the following authors: 
Spitzer et al.\ (1959a,b) \nocite{Spitzer59a} 
\nocite{Spitzer59b} performed thin film measurements on $\beta$-
and $\alpha$-SiC; Stephens (\cite{Stephens80}) measured on crystalline
$\beta$-SiC smokes;
Friedemann et al.\ (\cite{Friedemann81}) measured
two commercially available $\alpha$-SiC; 
Borghesi et al.\ (\cite{Borghesi85b}) investigated
three commercially produced $\alpha$- and one commercially produced
$\beta$-SiC; Papoular et al.\
(\cite{Papoular98}) measured two samples of $\beta$-SiC powders, one 
produced by laser pyrolysis and one which was commercially available;
Mutschke et al.\ (\cite{Mutschke99}) studied 16 different SiC powders
which were partly of commercial origin and partly laboratory products 
(8 $\alpha$-SiC and 8 $\beta$-SiC); Speck et al.\ (\cite{Speck99}) made thin
film measurements of $\alpha$- and $\beta$-SiC and Andersen et al.\ 
(\cite{Andersen99}) have measured the spectrum of meteoritic SiC.

One of the difficulties in interpreting laboratory data lies in 
disentangling the combination of several effects due to size, shape,
physical state (amorphous or crystalline), purity of the sample and
possible matrix effects if a matrix is used.
There is a general agreement that grain size and grain shape have a crucial
influence on the absorption feature of SiC particles. This is particularly
demonstrated by Papoular et al.\ (\cite{Papoular98}), Andersen et al.\
(\cite{Andersen99}) and Mutschke et al.\ (\cite{Mutschke99}).  
Papoular et al.\ (\cite{Papoular98}), 
Mutschke et al.\ (\cite{Mutschke99}) and Speck et al.\ (\cite{Speck99})
have shown 
that the matrix effect does not shift the resonance feature as a whole
as it was assumed by Friedemann et al.\ (\cite{Friedemann81}) and
Borghesi et al.\ (\cite{Borghesi85b}). While Papoular et al.\ 
(\cite{Papoular98}) and Mutschke et al.\ (\cite{Mutschke99}) find
that the profile is not shifted but altered, 
Speck et al.\ (\cite{Speck99}) state that the profile is not affected 
at all,  whether a matrix is used in the experimental set up or not.
The influence of purity of the laboratory samples was mainly studied by
Mutschke et al.\ (\cite{Mutschke99}). Another issue considered is the
effect of the crystal type. Silicon carbide shows pronounced polytypism 
which means that there exist a number of possible crystal types differing
in only one spatial direction. All these polytypes are variants of the
same basic structure and can therefore be divided into two basic
groups: $\alpha$-SiC (the hexagonal polytypes) and $\beta$-SiC
(the cubic polytype). It was found by Spitzer et al. 
(1959a,b)\nocite{Spitzer59a}\nocite{Spitzer59b}, Papoular et al.
(\cite{Papoular98}), Andersen et al.\ (\cite{Andersen99}) and
Mutschke et al.\ (\cite{Mutschke99}) that the crystal structure of
SiC cannot be determined from IR spectra, because there is no 
systematic dependence of the band profile on the crystal type.
In contrast, Borghesi et al.\ (\cite{Borghesi85b}) 
and Speck et al.\ (\cite{Speck99}) find the contrary result.

In this paper we have used the average value for bulk SiC reflectance
spectra of $\beta$-SiC as presented by Mutschke et al.\ (\cite{Mutschke99}) 
with
$\epsilon_{\infty} = 6.49$, $\omega_{\rm TO}$ = 795.4 cm$^{-1}$, 
$\omega_{\rm p}$ = 1423.3 cm$^{-1}$ and $\gamma$ = 10 
to calculate the optical constants $n$ and $k$,
using the one-oscillator model described in Mutschke et al.\ 
(\cite{Mutschke99}). The damping constant $\gamma$ is an
``ad hoc'' parameter, which in a perfect crystal reflects the anharmonicity of the
potential curve.  A damping constant of $\gamma = 10$ characterises 
crystals which are not structurally perfect but still far from amorphousness.

Since there is no systematic dependence of the band profile on the crystal
type in the data of Mutschke et al.\ (\cite{Mutschke99}), we could just as
well have used the data of one of their $\alpha$-SiC samples and
would have obtained a similar result. 

The optical constants $n$ and $k$ where used to calculate the extinction
efficiency for small spherical grains in the Rayleigh limit.
Spheres are not necessarily the best approximation for the grain shape of
SiC particles in C-rich AGB stars compared to 
e.g.\ a continuous distribution of ellipsoids (CDE) as introduced by
Bohren \& Huffman (\cite{Bohren83}).  The general
appearance of the feature as well as the peak position will depend on
the grain shape, however, common for all grain shapes of SiC are 
that the feature
will always fall between the transverse (TO) and the longitudinal (LO) 
optical phonon mode, so the difference will be that a spherical grain
shape will give rise to a sharper and narrower resonance than other grain
shape approximations.

\section{Dynamical models}\label{s:dynmod}

\subsection{Modelling method}\label{s:dmmeth}

To obtain the structure of the stellar atmos\-phere and circumstellar envelope
as a function of time
we solve the coupled system of radiation hydrodynamics and time-dependent 
dust formation (cf.\ H\"ofner et al.\ \cite{Hoefner95}, 
H\"ofner \& Dorfi \cite{Hoefner97}
and references therein). The gas dynamics is described by the equations of 
continuity, motion and energy, and the radiation field by the grey moment 
equations of the radiative transfer equation (including a variable Eddington 
factor). In contrast to the models presented in 
H\"ofner \& Dorfi (\cite{Hoefner97}) 
we use a Planck mean gas absorption coefficient based on detailed molecular 
data as described in H\"ofner et al.\ (\cite{Hoefner98}).
Dust formation is treated by the so-called moment method 
(Gail \& Sedlmayr \cite{Gail88}; Gauger et al.\ \cite{Gauger90}). 
We consider the formation of amorphous carbon 
in circumstellar envelopes of C-rich AGB stars.

\begin{figure}
 \centering 
 \leavevmode 
 \epsfxsize=1.00 
 \columnwidth
 \epsfbox{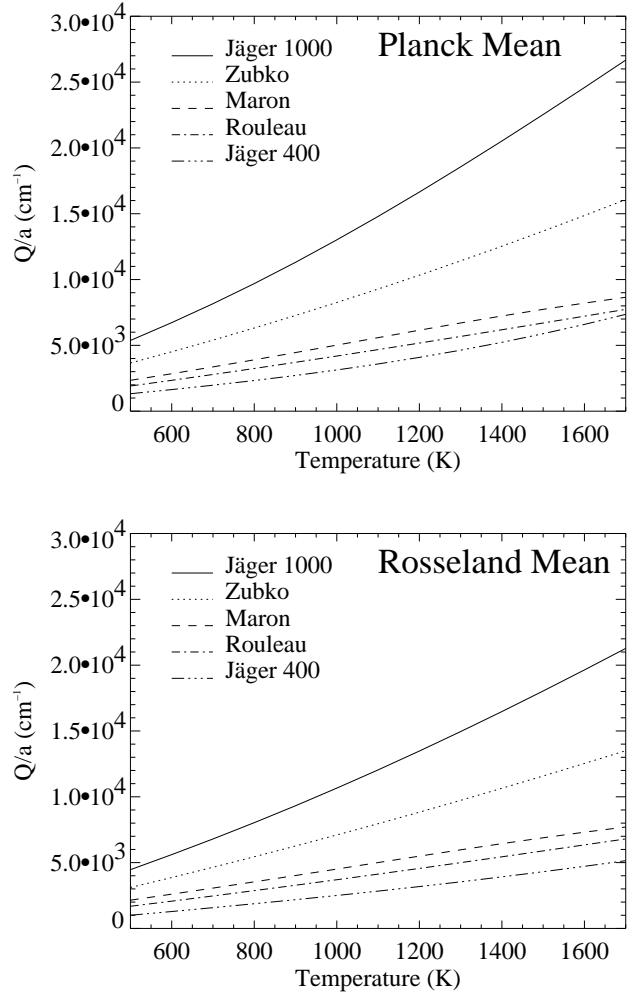}
  \caption[]{The Planck mean and the Rossland mean calculated from
the optical constants of amorphous carbon as derived from various experiments. 
See Table \ref{opacities} for 
details. The \PR\/ data would fall right on top of the \RO\/ data.}
  \label{mean}
\end{figure}

The dynamical calculations start with an initial model which 
represents the full hydrostatic limit case of the grey radiation hydrodynamics 
equations. It is determined by the following parameters:
luminosity $\Lstar$, mass $\Mstar$, effective temperature $\Tstar$
and the elemental abundances. We assume all elemental abundances to be solar
except the one of carbon which is specified by an additional parameter, i.e. 
the carbon-to-oxygen ratio $\CzuO$.
The stellar pulsation is simulated by a variable boundary (piston) which is
located beneath the stellar photosphere and is moving sinusoidally 
with a velocity amplitude $\dup$ and a period $P$.
Since the radiative flux is kept constant at the inner boundary throughout the 
cycle the luminosity there varies according to 
$L_{\rm in} (t) \propto R_{\rm in} (t) ^2$.

\subsection{Dust opacities}\label{s:dmdust}

The self-consistent modelling of circumstellar dust shells requires the 
knowledge of the extinction efficiency $Q_{\rm ext}$ 
of the grains, or rather of the quantity $Q_{\rm ext} / a$,
which is independent of $a$ 
in the small particle limit which is applicable in this context.
For the models of long period variables 
presented in H\"ofner \& Dorfi (\cite{Hoefner97}) and 
H\"ofner et al.\ (\cite{Hoefner98}) a fit formula for the Rosseland mean of
$Q_{\rm ext} / a$ derived from the optical constants of Maron (\cite{Maron90}) was
used.

One important point of this paper is to investigate the direct influence of
$Q_{\rm ext} / a$ on the structure and wind properties of the dynamical models.
Therefore we have computed Rosseland and Planck mean values of $Q_{\rm ext} / a$ 
(see Fig.\,\ref{mean}) 
based on various optical constants derived 
from laboratory experiments (see Sect.\,\ref{opaci} for details
about the samples).

For the dynamical calculations presented here we have selected the following 
data sets (see Table \ref{opacities} for a detailed specification):
\Jf\/ and \Jt\/ (representing the extreme cases),
\RO\/ (closest to the \MA\/ data used in earlier models but
extending to wavelengths below $1\,\mu$m) and \ZU.
The data of \PR\/ are almost identical to the data of \RO.

Figure \ref{mean} demonstrates that for a given data set the difference between
Planck and Rosseland means is relatively small. This is due to the fact
that amorphous carbon grains have a continuous opacity with a smooth 
wavelength dependence\footnote{ In contrast, the two means may differ
by orders of magnitude for gas opacities in case of molecular line blanketing.}.

\subsection{Wind properties}\label{s:dmwind}

All models discussed here are calculated with the same set of stellar 
parameters, i.e. $\Lstar = 13000 \,\Lsun$, $\Mstar = 1.0 \,\Msun$,
$\Tstar = 2700$\,K, $\CzuO = 1.4$, $P = 650$\,d, $\dup = 4$\,km/s,
corresponding to model P13C14U4 in H\"ofner et al.\ (\cite{Hoefner98}). 
The only 
difference between individual models is the adopted mean dust opacity.
Most models have been calculated with Rosseland mean dust opacities 
to allow us a direct comparison with earlier models based on Rosseland 
means derived from the \MA\/ data\footnote{Note however that
all models use Planck mean gas opacities as discussed in
Sect.\,\ref{s:dmmeth}.}.

\begin{table}
\caption{Comparison of modelling results for different dust opacity data:
         mass loss rate $\dot{M}$ (in $\msyr$), 
         mean velocity at the outer boundary $\uex$ (in km/s),
         mean degree of condensation at the outer boundary $\lra{\fcond}$;
         model parameters: $\Lstar = 13000 \,\Lsun$, $\Mstar = 1.0 \,\Msun$, 
         $\Tstar = 2700$\,K, $\CzuO = 1.4$, $P = 650$\,d, $\dup = 4$\,km/s;
         `R' denotes a Rosseland mean dust absorption coefficient, 
         `P' a Planck mean; for details see text.
         }\label{t:dynmod}
\begin{tabular}{|l|lllll|}
 \hline
  model & data & mean & $\dot{M}$         & $\uex$   & $\lra{\fcond}$ \\
 \hline \hline
  DJ1R      & \Jt & R  & $3.1 \cdot 10^{-6}$ & 14       & 0.17 \\
  DZUR      & \ZU & R  & $3.3 \cdot 10^{-6}$ & 13       & 0.21 \\
  DMAR$^*$  & \MA & R  & $2.9 \cdot 10^{-6}$ & 11       & 0.25 \\
  DROR      & \RO & R  & $2.7 \cdot 10^{-6}$ & 11       & 0.28 \\
  DJ4R      & \Jf & R  & $2.8 \cdot 10^{-6}$ & ~9       & 0.35 \\
 \hline \hline
  DJ1P      & \Jt & P  & $3.4 \cdot 10^{-6}$ & 14       & 0.15 \\
  DJ4P      & \Jf & P  & $2.9 \cdot 10^{-6}$ & 10       & 0.30 \\
 \hline
  \multicolumn{6}{l}{$^{*}$ model P13C14U4 of H\"ofner et al. (1998)} \\
\end{tabular}
\end{table}

Wind properties like the mass loss rate $\dot{M}$ or the time-averaged 
outflow velocity $\uex$ and degree of condensation $\lra{\fcond}$ are
direct results of the dynamical calculations.
The Rosseland mean models in Table \ref{t:dynmod} (first group) 
are listed in order of decreasing dust extinction efficiency. Both
$\uex$ and $\lra{\fcond}$
change significantly with the dust data used. $\lra{\fcond}$ increases
with decreasing dust extinction efficiency while $\uex$ decreases,
reflecting a lower optical
depth of the circumstellar dust shell (see also Sect.\,\ref{sed}).
The mass loss rates seem to show a weak overall trend but it is doubtful
whether the differences between ``neighbouring'' models in Table \ref{t:dynmod}
are significant. Since the mass loss rate varies strongly with time the
average values given in the table are more uncertain than the ones for 
the velocity and the degree of condensation which both do not show large 
variations with time.

The behaviour of the wind properties can be explained in the following 
way: The stellar parameters of the models presented here were chosen in 
such a way that the models fall into a domain where dust formation is 
efficient and the outflow can be easily driven by radiation pressure on 
dust (luminous, cool star, relatively high C/O ratio). In this case, the 
mass loss rate is essentially determined by the density in the dust 
formation zone which mainly depends on stellar and pulsation parameters 
(see e.g. H\"ofner \& Dorfi 1997). Therefore it is not surprising that the 
mass loss rates of the different models are quite similar. 

On the other hand, in a self-consistent model, the degree of condensation 
(dust-to-gas ratio) depends both on the thermodynamical conditions in the 
region where the dust is formed and on the specific grain opacity. The 
higher the mass absorption coefficient the faster the material is pushed 
out of the zone where efficient dust formation and grain growth is 
possible. Therefore the degree of condensation decreases with increasing 
dust absorption coefficients (i.e. higher radiative pressure) as grain 
growth is slowed down by dilution of the gas. Note that even in the model 
with the lowest dust absorption coefficient (DJ4R) the condensation of 
"free" carbon (i.e. all carbon not locked in CO) is far from complete 
($\lra{\fcond} < 1$).

For the two extreme cases (\Jt\/ and \Jf\/) we have also calculated models with
Planck mean dust opacities. As shown in Table \ref{t:dynmod} the wind
pro\-perties
of the corresponding Planck and Rosseland models (DJ1P/DJ1R and DJ4P/DJ4R) 
are very similar (if the differences are significant at all, see above).
The two Planck mean mo\-dels fit nicely into the dust extinction
efficiency sequence discussed before for the Rosseland mean models. 

As demonstrated in many earlier papers (e.g. Winters et al.\ \cite{Winters94}; 
H\"ofner \& Dorfi \cite{Hoefner97}) the dust formation in 
dynamical models is not necessarily periodic with 
the pulsation period $P$. In general the models are multi- or non-periodic
in the sense that the dust formation cycle is a more or less well defined
multiple of $P$. In the models discussed here, a new dust shell is formed 
about every 5-6 pulsation periods.

In contrast to the hotter inner regions, the atmospheric structure below 
about 1500\,K does not repeat after each pulsation cycle
but is more or less periodic on the dust formation time scales which 
span several pulsation cycles.
While it is easy to compare the time-averaged wind pro\-perties of the
models it is much more problematic to find comparable ``snapshots'' in 
different model sequences for the discussion of observable properties
presented in the next section. 
Therefore in all cases involving detailed radiative transfer on top of
given model structures we have decided to show statistical comparisons 
including several maximum and minimum phase models of each sequence.

\section{Spectral energy distributions and synthetic colours}\label{sed}

\subsection{Frequency-dependent radiative transfer}

The dynamical calculation yields the structure of the atmosphere and
circumstellar envelope (density, temperature, degree of condensation,
etc.) as a function of time. The time-independent radiative
  transfer equation is solved for each frequency separately along parallel rays
  to obtain spectral energy distributions (Windsteig et al.\
  \cite{Windsteig97} and references therein). 
The grey gas opacity (Planck mean) 
is taken directly from the dynamical models. The dust opacity is
calculated from the optical properties of amorphous carbon and 
in some cases SiC (see Sect.\,\ref{opaci} for the description of
the different dust data).

\subsection{Spectral energy distributions}

\begin{figure}
\centering 
 \leavevmode 
 \epsfxsize=1.00 
 \columnwidth
 \epsfbox{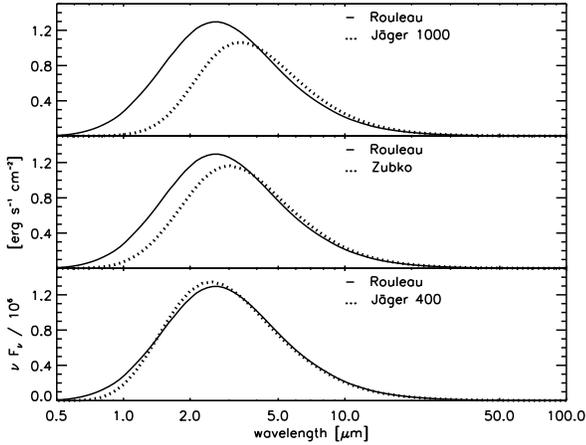}
\caption[]{{Spectral energy distributions of a minimum phase model of
  DROR where the spectra have been calculated with different
  dust data: we compare \RO\/ (consistent calculation) with \Jt\/
  (upper), with \ZU\/ (middle) and \Jf\/ (lower). For details see text.}} 
\label{incon}
\end{figure}

To investigate how the different dust data influence the resulting
spectral energy distributions (SEDs), spectra were calculated using 
the optical constants of amorphous carbon of \RO, 
\ZU, \Jf\/ and \Jt\/ on top of a fixed atmospheric structure. 
Two different kinds of SEDs were calculated; 
(1) fully consistent ones where the same amorphous carbon data were
used in the dynamical model and in the SED calculations and 
(2) ``inconsistent'' ones where we used different dust opacity data
for the detailed radiative transfer on top of the same dynamical model
structure (fixed spatial distribution of density, temperature, degree
of condensation). The latter spectra enable us
to distinguish between the effect of the various dust data in the radiative
transfer calculation and the effect on the model structure.

Figure \ref{incon} shows the result for the
SEDs based on a mini\-mum phase model of the DROR model sequence. The full
line always denotes the SED of the consistent model where the
\RO\/ data were used for the underlying dynamical
model as well as for the calculation of the spectrum. 
In the upper panel the (inconsistent) \Jt\/ 
spectrum (dotted) calculated on top of the same \RO\/ model is
shown in addition to the consistent \RO\/ spectrum. The
middle panel shows the same for \ZU\/ and the lower panel for \Jf.
The effects of the different dust data for a given structure compared to a 
consistent model using the \RO\/ data can be summarised as follows:

\begin{itemize}
\item \Jt: the spectrum has a lower flux in the short wavelength
 region (0.5 to 5\,$\mu$m) and the maximum 
 at longer wavelengths. The lower flux level in this region is due to
 the fact that $Q_{\rm ext}/a$ for the \Jt\/ data is higher than for the \RO\/ data, 
 therefore we have a higher total dust opacity which results in less flux 
 coming out. The shift of the maximum is also due to the higher dust opacity 
 in the \Jt\/ case.
\item \ZU: the spectrum has a lower flux level in the short wavelength
 region and the maximum 
 is shifted to longer wavelengths, but not as far as the \Jt\/ spectrum.
 This is due to the fact that \Jt\/ is a more "graphite-like"
 amorphous carbon dust than the \ZU\/ material. 
\item \Jf: the spectrum has a comparable flux all over the spectrum and the
 maximum at slightly shorter wavelengths.
 The slightly higher flux level around the maximum results from the
 lower $Q_{\rm ext}/a$ of the \Jf\/ data which is due to its more
 ``diamond-like'' nature compared to the \RO\/ data.
 In the wavelength region where the maxima of the 
 spectra lie, the two data sets are very similar, therefore the maxima of 
 the SEDs do not differ much in wavelength.
\end{itemize}

Note that for the ``inconsistent'' SEDs the total flux may differ from
the value of the consistent models.

\subsection{Spectral energy distributions including SiC}

\begin{figure}
\centering 
 \leavevmode 
 \epsfxsize=1.00 
 \columnwidth
 \epsfbox{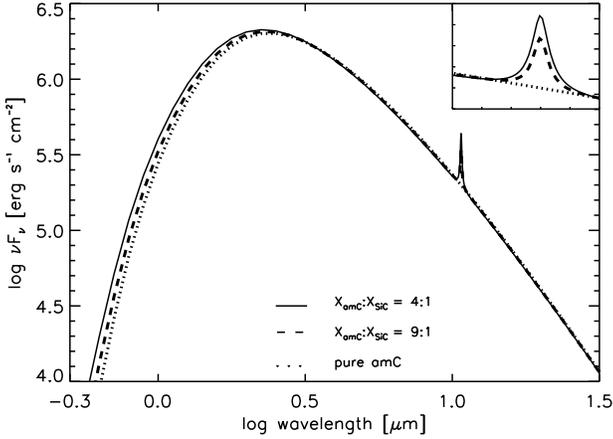}
\caption[]{Spectral energy distributions of a minimum phase model of
  DJ1R where the spectra have been calculated with amorphous
  carbon alone (dotted), with a ratio X$_{\rm amC}$:X$_{\rm SiC}$ of
  9:1 (dashed) and a ratio of X$_{\rm amC}$:X$_{\rm SiC}$ of 4:1 (solid).} 
\label{sic}
\end{figure}

The analysis of mid-IR carbon star spectra indicates that SiC is the
best candidate to reproduce the observations around the 11~$\mu$m
region. We have therefore considered SiC as an additional dust
component (see Sect.\,\ref{SiC} for details). 
The formation of SiC is not included in the
self-consistent model calculations because (1) little is known about
the condensation process and (2) because we do not expect that SiC
will have a significant influence on the model structures. We use
either Planck or Rosseland means for the model computations and SiC
will contribute only in a very narrow wavelength region with small
amounts to these mean opacities compared to amorphous carbon.
 
The effect of SiC as dust component is described in a qualitative
manner. The dust opacity $\kd$ for each wavelength is calculated from 
\begin{center}
$\kd = \frac{\kappa_{\rm amC}X_{\rm amC} + \kappa_{\rm SiC}X_{\rm SiC}}
{X_{\rm amC} + X_{\rm SiC}}$
\end{center}
\noindent
where X$_{\rm i}$ are the fractional parts of amorphous carbon and SiC,
respectively, and where $\kappa_{\rm amC}$ and $\kappa_{\rm SiC}$ are the opacities of carbon and SiC.
Figure \ref{sic} shows how a mixture of
dust grains consisting of amorphous carbon (\Jt\/ data) and SiC
modifies the SED around 11.3 $\mu$m. 
Two different ratios of X$_{\rm amC}$ : X$_{\rm SiC}$, 
4:1 and 9:1 were adopted. The
higher the amount of SiC, the stronger is the 11.3~$\mu$m feature (see
inset of Fig. \ref{sic}).
Another choice of grain shape than spherical for the SiC particles, would
result in a broader and weaker feature.

\subsection{Synthetic colours}

\begin{figure*}
\centering 
 \leavevmode 
 \epsfxsize=1.50 
 \columnwidth
 \epsfbox{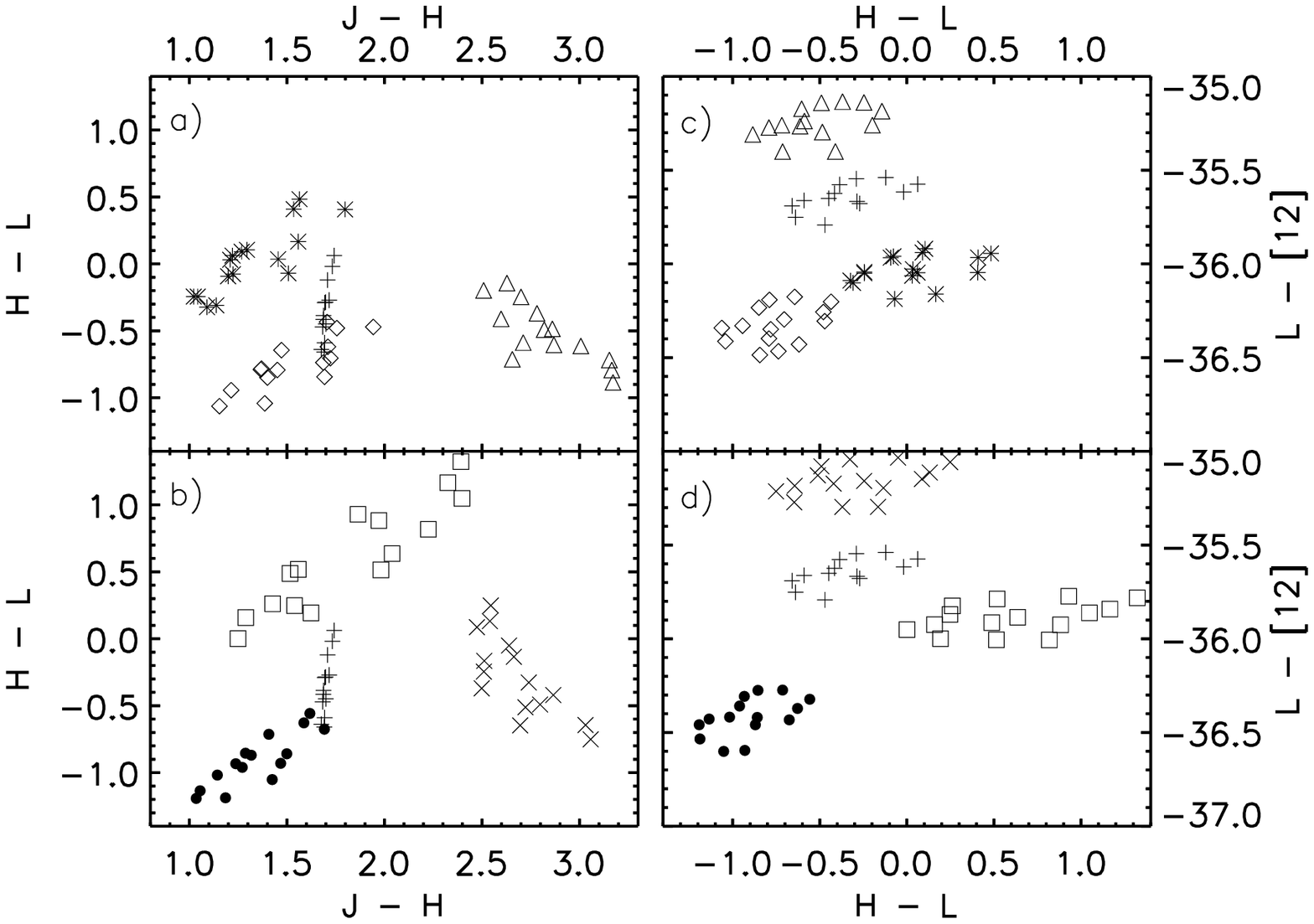}
\caption[]{Synthetic colours:
  Upper panels: 
  left: (J$-$H) vs. (H$-$L) for the maxima and minima of consistent models; 
    crosses denote the \RO\/ data (DROR), 
    asterisks the \Jt\/ data (DJ1R), 
    diamonds the \Jf\/ data (DJ4R) and 
    triangles the \ZU\/ data (DZUR); 
  right: (H$-$L) vs. (L$-$[12]) for the same models (same symbols). 
  Lower panels: ``inconsistent'' colours, all based on model DROR,
  in comparison to the consistent \RO\/ colours (crosses): 
    circles represent spectra calculated with the \Jf\/ data,
    squares denote the \Jt\/ data and
    x the \ZU\/ dust data. 
  This plot shows that the influence
  of the different dust data used in the radiative transfer calculation
  is much stronger than the effect of the model structures.}
\label{colours}
\end{figure*}

For a comparison of the consistent spectra (model structure and spectra
computed with the same dust data) we have calculated synthetic
J, H and L colours as well as the IRAS 12~$\mu$m colour. 
In a (J$-$H) versus (H$-$L) dia\-gram (Fig.\,\ref{colours}a) the
models based on different amorphous dust data fall into distinct regions. 

The models calculated with the \Jt\/ dust data have the
reddest colours in (H$-$L). The \Jf\/ colours are the bluest, while
\RO\/ and \ZU\/ lie in between. In (J$-$H) models with the \ZU\/
data are reddest and the others do not differ much.
The reason for the diffe\-rent slopes of \RO\/ and \ZU\/
compared to both of the {\it J\"ager} data sets is that in these cases the
maxima of the SEDs are changing between the J and the H filter
depen\-ding on the phase. The maxima of the SEDs resulting from the
\Jt\/ model structures are always at longer wavelengths
and the ones of the \Jf\/ structures lie mainly in one filter.
From Fig.\,\ref{colours}b, which shows the ``inconsistent'' colours
based on model DROR (structure was calculated with the \RO\/ data
and the spectra with other dust data) in addition to the consistent
\RO\/ colours, it is clear
that the influence of the different dust data used in the radiative
transfer calculation is much stronger than the effect of 
the underlying hydrodynamic model structure. 

\begin{figure}
\centering 
 \leavevmode 
 \epsfxsize=0.90 
 \columnwidth
 \epsfbox{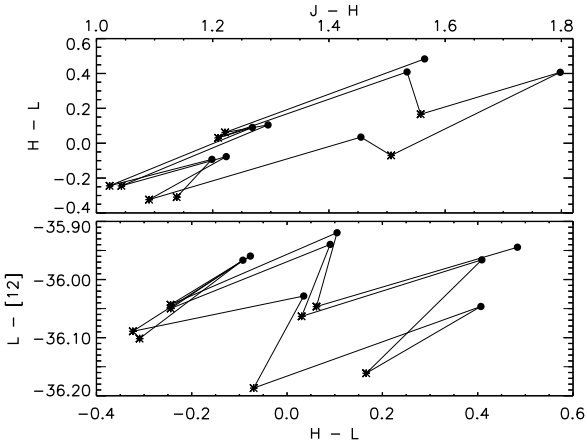}
\caption[]{Upper panel: (J$-$H) vs. (H$-$L) for maxima (asterisks) and
  minima (circles) of the DJ1R model. When connecting the points
  following each other one can see that the result is a spiral, the
  minima are always redder than the following maxima. The colours are
  strongly related to the formation of a new dust shell which takes
  place every 5 to 6 pulsation cycles. After this time scale the
  colours match very closely the ones during the formation of the prior
  dust shell. Lower panel: Same as above, only (H$-$L) vs. (L$-$[12])
  is shown.}
\label{loop}
\end{figure}

Note that in Fig.\,\ref{colours} only maximum and minimum phases are
shown. Other phases would fill in the gaps between successive
extremes. The colours are strongly related to the formation of a new
dust shell which takes place every 5 to 6 pulsation cycles (see
Sect.\,\ref{s:dmwind}). After this time scale the colours match very closely
the ones of the preceding dust formation cycle as shown in Fig.\,\ref{loop}.
When connecting the succeeding points it can be seen that they form 
a spiral. The minima (circles) are always redder than the following
maxima (asterisks).

To investigate also the mid-IR properties of the models we calculated
the 12~$\mu$m colour. A (L$-$[12]) vs. (H$-$L) diagram
(Fig.\,\ref{colours}c) shows, that again the consistent colours fall
into distinct regions. The sequence in
(L$-$[12]) (\ZU\/ - \RO\/ -  \Jt\/ -  \Jf) is a
sequence of decreasing optical depths. 
Table \ref{tau} lists
the mean dust optical depths for a few selected wavelengths.
In Fig.\,\ref{colours}d the inconsistent colours based on the model 
structure of DROR are shown for comparison.

\begin{table}
\caption{Mean dust optical depths at a few selected wavelengths}\label{tau}
\begin{tabular}{|l|lllll|}
 \hline
  optical depth & DJ1R& DZUR & DROR & DJ4R & \\
 \hline \hline
  $\tau_{d,1\mu}$ & 1.25& 1.1 & 1.0 & 1.15 & \\
  $\tau_{d,2.2\mu}$ & 0.45 & 0.37 & 0.37 & 0.23 & \\
  $\tau_{d,10\mu}$ & 0.06 &  0.08 & 0.07 & 0.04 & \\
 \hline
\end{tabular}
\end{table}
 
From Fig.\,\ref{colours} we can infer that the influence of the different
amorphous carbon dust data used in the radiative transfer calculation
is much stronger than the effect of the model structures. The colours
resulting from the same dust data fall approximately into the same
region of a two-colour-diagram, whether they are calculated on top of
the corresponding model structure (upper panel) or a fixed model
sequence (lower panel). This applies for all data sets. 

In addition we
computed synthetic colours for the \MA\/ and the
\PR\/ data, but they are not shown in Fig. \ref{colours}
to avoid overlaps in the plot. The \PR\/ colours fall into the
same region as the \ZU\/ colours as one would expect because 
the $Q_{\rm ext}/a$  are very similar. The same applies for the \MA\/ and 
the \RO\/ data. Only the J flux differs between \MA\/ and \RO, 
the reason being that the \MA\/ data do not extend below 1~$\mu$m and 
therefore the corresponding contribution to the J filter is lacking.

\section{Summary and conclusions}

Carbon bearing grains are expected to form in the outflows of 
C-rich AGB stars. The two most common types of carbon grains in
these stars are expected to be amorphous carbon and SiC grains.
We have investigated the direct influence of different dust
optical properties on the wind characteristics and the resulting
observable properties of the dynamical models.

The term amorphous carbon covers a wide variety of material properties from
``diamond-like'' to ``graphite-like''. 
We have used $n$ and $k$ data of Maron (\cite{Maron90}),
Rouleau \& Martin (\cite{Rouleau91}), Preibisch et al.\ (\cite{Preibisch93}), 
Zubko et al.\ (\cite{Zubko96}) and J\"ager et al.\ (\cite{Jager98}) to
investigate the influence of different types of amorphous carbon, 
on the structure and the wind properties of dynamical
models. The Rosseland and Planck mean values of Q$_{ext}$/a 
used in the model computations were calculated 
from the Rayleigh approximation for spheres.  The difference between the
Planck and the Rosseland means is relatively small for the amorphous carbon
data because the grains have a continuous opacity with a smooth wavelength
dependence.

In our test models,
both the outflow velocity $\uex$ and degree of condensation $\lra{\fcond}$ 
change significantly with the dust data used. $\lra{\fcond}$ increases with
decreasing dust extinction efficiency while $\uex$ decreases, reflecting
a lower optical depth of the circumstellar dust shell.  The mass loss
rate is, however, not significantly influenced by the use of different
dust data.  

On top of the structures resulting from the dynamic calculations
we have performed detailed radiative transfer calculations to obtain the
spectral energy distribution of the circumstellar dust shells.
Regarding infrared colours, 
the influence of the different dust data used in the radiative transfer
calculation is much stronger than the effect of the underlying hydrodynamic
model structure.  
However, this should not be used as an excuse for fitting observations
by arbitrarily choosing the optical properties of the dust grains 
for a given model structure. In a consistent model the dynamical
properties (e.g.\ outflow velocities) and the optical appearance of
the circumstellar envelope are related in a complex way.

The influence of including SiC grains is that the 11.3~$\mu$m
feature appears in the spectral energy distribution of the models. How
much SiC should be ``mixed'' into a model to reproduce the 11.3 $\mu$m feature
observed (e.g.\ class 4 in Goebel \cite{Goebel95}) will very much depend on the
assumptions which are made about the size and shape of the SiC grains which
enter into the model (Papoular et al.\ \cite{Papoular98}; Andersen et al.\
\cite{Andersen99}; Mutschke et al.\ \cite{Mutschke99}).

\begin{acknowledgements}

This work was supported by the Austrian Science Fund (FWF, project
number S7305-AST) and the Austrian Academy of Sciences (RL acknowledges a
``Doktoranden\-stipendium''). 
We thank E.A Dorfi (IfA Vienna) for inspiring discussions.

\end{acknowledgements}

\end{document}